\documentclass[prl,aps,twocolumn,floats,superscriptaddress]{revtex4}
\usepackage{amsmath}
\usepackage{graphicx}
\usepackage{epsfig}
\newlength{\upit}\upit=0.1truein

\newcommand{\ltappr}{{{\lower4pt\hbox{$<$} } \atop \widetilde{ \ \ \ }}}
\newlength{\bxwidth}\bxwidth=1.5 truein

\begin{document}
\newcommand{\dg}{^{\dagger }}
\newcommand{\si}{\sigma}
\newcommand{\rarrow}{\rightarrow}
\def\fig#1#2{\includegraphics[height=#1]{#2}}
\def\figx#1#2{\includegraphics[width=#1]{#2}}
\newlength{\figwidth}
\newlength{\shift}
\shift=0.4cm
\newcommand{\fg}[3]
{
\begin{figure}[ht]
\vspace*{-0cm}
\[
\includegraphics[width=\figwidth]{#1}
\]
\vspace*{\shift}
\caption{\label{#2}
\small
#3
}
\end{figure}}

\newcommand \bea {\begin{eqnarray} }
\newcommand \eea {\end{eqnarray}}



\title{Phase diagram of the $t-t^{\prime}-U$ chain at half filling}

\author{M.E. Torio$^1$, A.A. Aligia$^2$, and H.A. Ceccatto$^1$}
\affiliation{${^{1}}$ Instituto de F\'{i}sica Rosario, Consejo
Nacional de Investigaciones Cient\'{i}ficas y T\'ecnicas and
Universidad Nacional de Rosario, Boulevard
27 de Febrero 210 Bis, 2000 Rosario, Argentina\\
${^{2}}$ Centro At\'{o}mico Bariloche and Instituto Balseiro,
Comisi\'on Nacional de Energ\'{\i}a At\'{o}mica, 8400 Bariloche,
Argentina}

\date{\today}

\begin{abstract}
We study the half filled Hubbard chain including next-nearest-neighbor
hopping $t^{\prime }$. The model has three phases: one insulating phase with
dominant spin-density-wave correlations at large distances (SDWI), another
phase with dominant spin-dimer correlations or dimerized insulator (DI), and
a third one in which long distance correlations indicate singlet
superconductivity (SS). The boundaries of the SDWI are accurately determined
numerically through a crossing of excited energy levels equivalent to the
jump in the spin Berry phase. The DI-SS boundary is studied using several
indicators like correlation exponent $K_{\rho }$, Drude weight $D_{c}$,
localization parameter $z_{L}$ and charge gap $\Delta _{c}$.
\end{abstract}


\maketitle

\section{Introduction}

Superconductivity in low dimensional systems has been a subject of
interest in the last years. The physics of the negative-$U$
Hubbard model, \cite{negu} as well as that of the
ladder,\cite{noac,dagr} suggest that the opening of a spin gap
$\Delta _{s}$ is a key ingredient to superconductivity. In fact, a
non zero $\Delta _{s}$ suggests the existence of bound pairs of
electrons (like Cooper pairs), which if mobile can lead to a SS
state. The indications of resonance valence bond superconducting
states in effective low energy models for 2D cuprates, like the
generalized $t-J$ \cite{and,rvb} or the Hubbard model with
correlated hopping,\cite{ed} are also consistent with this
picture. Closely related with the latter system is the Hubbard
model with an extra nearest-neighbor exchange $J$,\cite{bos} for
which also strong signals of superconductivity have been
found.\cite{jeck,daul0}

The half-filled Hubbard model with additional
next-nearest-neighbor hopping $t^{\prime }$ in one dimension is
appealing because, for weak $U$, one expects the existence of a
metallic phase when $t^{\prime }>0.5t,$\cite{fab,note} and, from
previous studies of the frustrated Heisenberg chain,\cite
{nom,egg} one expects the opening of a spin gap for large $U$ and
$t^{\prime }/t >\sqrt{R_{c}}$ with $R_{c}=0.241167$.\cite{egg} In
addition, this model might be valid as an effective model for
several one dimensional cuprates.\cite{rosner,miz} The
corresponding Hamiltonian is:
\begin{eqnarray}
H&=&-t\sum_{i\sigma }(c_{i+1\sigma }^{\dagger }c_{i\sigma }+
\mathrm{H.c.})-t^{\prime }\sum_{i\sigma }(c_{i+2\sigma }^{\dagger
} c_{i\sigma }+\mathrm{H.c.}) \nonumber \\
&&+U\sum_{i}n_{i\uparrow
}n_{i\downarrow }.  \label{h}
\end{eqnarray}
\indent Using weak coupling renormalization group, Fabrizio
obtained that a spin gap opens for $t^{\prime }>0.5t$, and that
for $\Delta _{s}\neq 0$ SS correlations dominate in the metallic
phase while spin dimer correlations are the largest at long
distances in the insulating phase.\cite{fab} Later studies using
density matrix renormalization group (DMRG) sketched an
approximate phase diagram,\cite{daul} confirmed the SS character
in the metallic phase,\cite{kuro} and determined some points where
approximately the charge transition SS-DI takes
place.\cite{kuro,aebi}

In this work we present further evidence of the dominance of SS
correlations in the metallic phase by calculating the correlation
exponent $K_{\rho }$. We also obtain more information on the
metal-insulator transition using numerical calculations of the
Drude weight $D_{c}$, localization indicator $z_{L}$
\cite{resta,ort1,ort2} and charge gap $\Delta _{c}$. Furthermore,
we determine the opening of the spin gap $\Delta _{s}$ using the
crossing of singlet and triplet excitations,\cite {nom,nak} which
is equivalent to the jump of the spin Berry phase.\cite{epl} This
method has been shown before to lead to very accurate results for
this transition, even if the size $L$ of the systems considered is
less than 13.\cite{bos,nak,gag,hal,nak2,topo,pola,ih} As an
example, the resulting opening of the spin gap for the Hubbard
model with correlated hopping \cite{bos,topo} coincides with field
theory results in the weak coupling limit,\cite {bos,japar} and
with exact results when the number of doubly occupied sites is
conserved.\cite{afq} Moreover, this method allowed us to determine
the complete boundary of the SDWI phase.

In Section II we present the results for the spin transition and
briefly describe the method used. In Section III, we discuss our
results for the correlation exponent $K_{\rho }$. In Section IV we
explain briefly the meaning of the localization indicator $z_{L}$,
present results for it, for the Drude weight $D_{c}$, and for the
charge gap $\Delta _{c}$, and use this information to draw a
tentative a phase diagram. Section V contains a summary and
discussion.

\section{The spin transition}

This transition can be determined by a topological transition (a
jump in the spin Berry phase), which is equivalent to a suitable
crossing of excited energy levels. The basic idea of this crossing
is that, in conformal invariant systems (corresponding to fixed
points in weak coupling bosonization theory plus renormalization
group), the lowest excitation energy determines the dominant
correlations at large distances.\cite{nom,nak} On the other hand,
the topological transitions denote rearrangements of charge or
spin which have a very transparent meaning in the strong coupling
limit,\cite{gag,topo,ih} and are expected to be valid by
continuity also for weak coupling. For SU(2) invariant systems
both criteria coincide for the spin transition.\cite{epl}

\begin{figure}[tbd]
\includegraphics[width=\columnwidth]{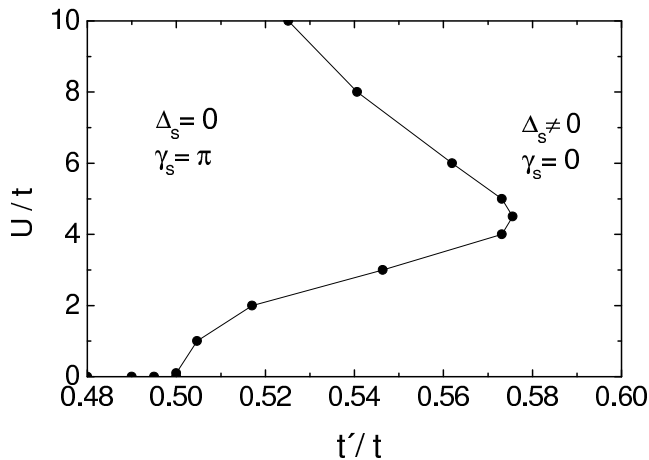}
\vskip -0.2cm
 \caption{\scriptsize Topological sectors of the spin Berry phase
in the $t^{\prime }$, $U$ plane ($t=1$). $\Delta _{s}=0$ for
$\gamma _{s}=\pi $ and $\Delta _{s}\neq 0$ for $\gamma _{s}=0$.
}\label{diagfastp}
\end{figure}

The spin Berry phase $\gamma _{s}$ is the phase captured by the
ground state using twisted boundary conditions --$e^{i\Phi }$ for
spin up and $e^{-i\Phi }$ for spin down-- in the cycle $0\le \Phi
\le 2\pi $.\cite{epl} In gauge invariant form, it can be
numerically calculated as:
\begin{eqnarray}
\gamma _{s} &=&-\lim_{N\rightarrow \infty }\{\text{Im}[\ln (\Pi
_{i=0}^{N-2}\langle g(\Phi _{i},-\Phi _{i})|g(\Phi _{i+1},-\Phi
_{i+1})\rangle  \nonumber \\
&&\langle g(\Phi _{N-1},-\Phi _{N-1})|g(2\pi )\rangle )]\},  \label{gs}
\end{eqnarray}
where $|g(\Phi _{\uparrow },\ \Phi _{\downarrow })\rangle $\ is
the ground state of $\tilde{H}(\Phi _{\uparrow },\ \Phi
_{\downarrow })$, and $|g(2\pi )\rangle =\exp [i{\frac{2\pi
}{L}}\sum_{j}j(n_{j\uparrow }-n_{j\downarrow })]|g(0,0)\rangle $.
The Hamiltonian $\widetilde{H}$ differs from $H$ in that the
hopping term has the form $-t\sum_{i\sigma }
(\widetilde{c}_{i+1\sigma }^{\dagger }\widetilde{c}_{i\sigma
}e^{i\phi _{\sigma }/L}+\mathrm{H.c.}).$

The Berry phase $\gamma _{s}$ can take only two values: 0 or $\pi$
(modulo $2\pi $). The jump between these values is related to the
change in the difference between electric polarization from spin
up and down by $\Delta \gamma _{s}=(2\pi /e)\Delta (P_{\uparrow
}-P_{\downarrow })$.\cite{epl} The position of the jump of $\gamma
_{s}$ coincides with the singlet-triplet crossing of the energy
levels: the crossing of a singlet even under inversion with an odd
triplet, with $K=0$. The boundary conditions are periodic
(antiperiodic) for system lengths $L=4n$ ($4n+2 $), with $n$
integer (they are \emph{opposite} to the so called closed shell
ones). The same crossing is the one which determines the opening
of the spin gap (the Kosterlitz-Thouless transition in the spin
sector) according to conformal field theory, after bosonizing the
problem and using renormalization group.\cite{nak} This theory
predicts a $1/L^{2}$ scaling of the transition point.

We have calculated the transition for $L=8$, 12 and 16. The
restriction to $L$ values multiple of four is due to the fact that
other system sizes lead to frustration of the spin-density waves
(or classical antiferromagnetic ordering) expected for large $U$.
In particular, for $t=0$ the even and odd sites are decoupled and,
to avoid frustration of the AF classical order in each subsystem,
$L/2$ should be even. We have fitted the resulting three points
with a straight line $a+b/L^{2}$ in $1/L^{2}$. The transition
point in the thermodynamic limit is determined by $a$ and its
error gives an indication of the fit quality.



In Fig. \ref{diagfastp} we show the results for the spin
transition for $U\leq 10$ (we have chosen $t=1$ as the unit of
energy). For $t^{\prime }<0.5$ we obtain that the transition takes
place at $U=0$ for all $L$. For positives values of $U$, $\gamma
_{s}=\pi $ and $\Delta _{s}=0$, while for negative $U$, $\gamma
_{s}=0$ and $\Delta _{s}\neq 0$, as in the ordinary Hubbard
model.\cite{topo} At the point $U=0$, $t^{\prime }=0.5$, the spin
transition line deviates $90^{\circ }$, and for positive $U$ its
shape is like a nose, with a maximum value of $t^{\prime
}=0.564\pm 0.001$ taking place near $U=4.5$. The errors of the
points shown in Fig. \ref{diagfastp} are less than 0.003.

\begin{figure}[tbd]
\includegraphics[width=\columnwidth]{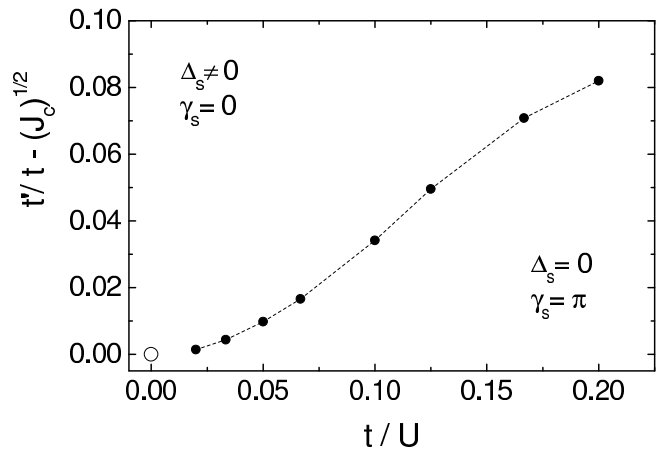}
\vskip -0.2cm \caption{\scriptsize Same as Fig. 1 in the $1/U$,
$t^{\prime }$ plane. The open symbol corresponds to the known
result $t^{\prime }=\sqrt{R_{c}}=0.4911$ for $U\rightarrow \infty
$.}\label{tpvsu}
\end{figure}

Results for larger values of $U$ are shown in Fig. \ref{tpvsu}.
For $t,t^{\prime }\ll U$, to leading order in $1/U$ the model is
equivalent to the Heisenberg chain including next nearest-neighbor
exchange:
\begin{equation}
H=\sum_{i}(J {\bf S}_{i} \cdot {\bf S}_{i+1} +J^{\prime }{\bf
S}_{i}\cdot {\bf S}_{i+2})  \label{hheis}
\end{equation}
with $J=4t^{2}/U$, $J^{\prime }=4t^{\prime }{}^{2}/U$. Previous
studies of this model using the method of crossings of excitation
levels plus a detailed knowledge of the size dependence have
established the critical value $R_{c}=(J^{\prime }/J)_{c}\simeq
0.241167$ at the transition.\cite{egg} Then, for $U\rightarrow
\infty $, $\Delta _{s}$ should open at $ t_{c}^{\prime
}=\sqrt{R_{c}}$. Our results are consistent with this fact. A fit
of our transition points at $U=15$, 20, 30, and 50 with the
parabola $t_{c}^{\prime }-\sqrt{R_{c}}=a+b/U+c/U^{2}$ gives
$c=3.6$ and $a=b=0$ within our numerical error for $t_{c}^{\prime
}$ (less than $5\times 10^{-4}$). The fact that $b=0$ is also
consistent with the strong coupling expansion: $b$ is proportional
to the first correction to Eq. (\ref{hheis}), which is of order
$t^{2}t^{\prime }/U^{2}$. However, the electron-hole
transformation $c_{i\sigma }^{\dagger }\leftrightarrow c_{i\sigma
}$ changes the sign of both hoppings but should leave invariant
the spin gap at half filling. Then, all the corrections to $H$ of
order $1/U^{2n}$ with $n$ integer should either vanish or be
irrelevant for the spin gap since they change sign under
electron-hole symmetry.

\section{Correlation exponent in the metallic phase}

In the metallic phase, the analysis based on bosonization and
renormalization group\cite{fab} shows that there is only one
gapless charge model and the spin modes are gapped ($\Delta
_{s}\neq 0$). This analysis is valid at weak coupling and requires
that $t^{\prime }{}>0.5$.\cite{note} In this case, there are four
Fermi points at wave vectors $\pm k_{F1}$ and $\pm k_{F2}$. The
decay of the different correlation functions at large distances is
determined by the correlation exponent $K_{\rho }$. Depending on
the value of this exponent, the slowest asymptotic decay
corresponds either to spin dimer-dimer correlation functions or to
superconducting ones. At half filling the former have the
following decay:
\begin{eqnarray}
\chi _{DW}(x)&=&\langle O_{DW}(x)O_{DW}(0)\rangle \sim \frac{\cos
[2(k_{F2}-k_{F1})x]}{x^{2K_{\rho }}} \nonumber \\
&=&\frac{\cos (\pi x)}{x^{2K_{\rho }}},
\label{corrpares}
\end{eqnarray}
where
\begin{equation}
O_{DW}(x)=S^{+}(x)S^{-}(x+a)+S^{+}(x+a)S^{-}(x).  \label{odw}
\end{equation}
For the dominant superconducting correlations one has
\begin{equation}
\chi _{SC}(x)=\langle \Delta (x)\Delta ^{\dag }(0)\rangle \sim \frac{1}{%
x^{1/2K_{\rho }}},  \label{corrsc}
\end{equation}
where
\begin{eqnarray}
\Delta (x)=\sum_{p=\pm }\psi _{pk_{F1}\uparrow }(x):\psi
_{-pk_{F1}\downarrow }(x) \nonumber \\
-\psi _{pk_{F2}\uparrow }(x):\psi_{-pk_{F2}\downarrow }(x).  \label{deltasc}
\end{eqnarray}

\begin{figure}[tbd]
\includegraphics[width=\columnwidth]{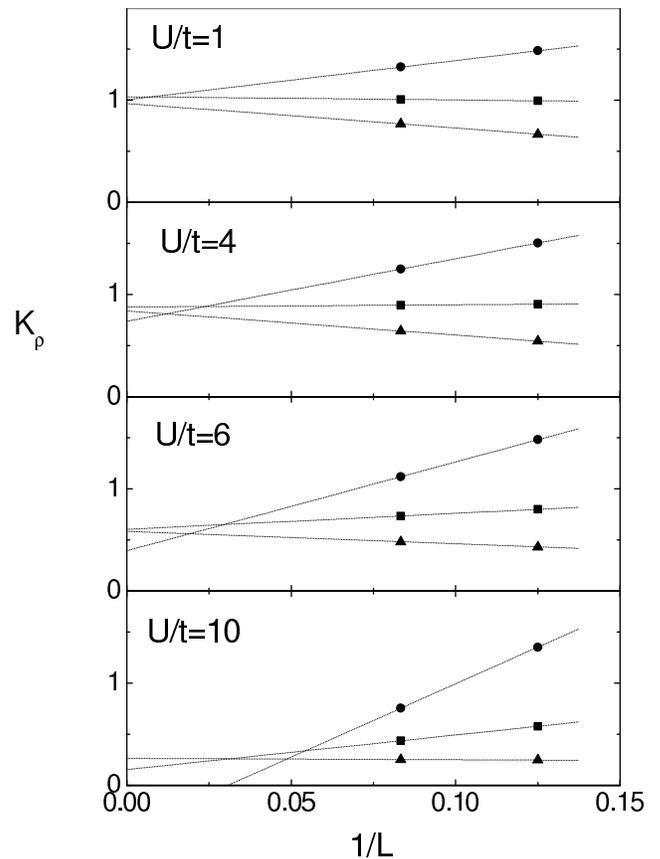}
\vskip -0.2cm \caption{\scriptsize Correlation exponent $K_{\rho
i}$ defined by Eqs. (\ref{krho1y2}) as a function of $1/L$, for
several values of $U$ and $t^{\prime }=2.5$. Triangles $K_{\rho
1}$, circles $K_{\rho 2}$, squares $K_{\rho 3}$.}\label{figkrho}
\end{figure}

An analysis similar to that of Schulz\cite{schulz} extended to
this case with 4 Fermi points leads to the following relations
involving the correlation exponent $K_{\rho }$:
\begin{equation}
\frac{\pi v_{\rho }}{4K_{\rho 1}}=\frac{1}{\kappa n^{2}},\;\;\;D_{c}=\frac{%
2v_{\rho }K_{\rho 2}}{\pi }.  \label{krho1y2}
\end{equation}
Here, the subscript 1 or 2 in $K_{\rho }$ is due to the fact that,
in finite systems, when $K_{\rho }$ is calculated using these
expressions the results differ. In the above equation $v_{\rho }$
is the charge velocity, $\kappa $ is the compressibility and
$D_{c}$ is the Drude weight. In finite systems these quantities
can be calculated as:
\begin{equation}
E(K=2\pi /L)-E(K=0)=\frac{2\pi }{L}v_{\rho }.  \label{vcarga}
\end{equation}
\begin{equation}
\frac{1}{\kappa n^{2}}=L\frac{E(N+2)+E(N-2)-2E(N)}{4}.  \label{compresnumer}
\end{equation}
\begin{equation}
D_{c}=\frac{L}{2}\frac{\partial ^{2}E_{0}(L,\Phi )}{\partial \Phi ^{2}}%
|_{\Phi = \Phi _{0}},  \label{drude}
\end{equation}
where $E(K)$ is the energy of the lowest singlet with total wave
vector $K$, $E(L,N)$ is the ground state energy for length $L$ and
$N$ particles, and $E_{0}(L,\Phi )$ is $E(L,L)$ in presence of a
flux $\Phi h/(2\pi e)$ threading the ring. Unless otherwise
stated, the boundary conditions correspond to the ``closed shell''
ones (those which lead to the minimum energy): $\Phi =0$ $(\pi )$
for $N=2n$ with $N$ odd (even).

The above equations lead to two independent ways of calculating
$K_{\rho }$. A third possibility uses the geometric mean $K_{\rho
3}=\sqrt{K_{\rho 1}K_{\rho 2}}$. In $K_{\rho 3}$ the charge
velocity does not enter, and usually it leads to a more accurate
result. The consistency of the three results in the thermodynamic
limit is a test on the validity of the Luttinger liquid
description, with gapless charge excitations. In fact, when the
$K_{\rho i}$ extrapolated to the thermodynamic limit start to
deviate from each other one has a strong indication of a
metal-insulator transition.\cite{physc} For example, in the model
with correlated hopping, the point where the $K_{\rho i}$ start to
deviate agrees with the jump in the charge Berry phase which
corresponds to this transition.\cite{topo,physc}

In Fig. \ref{figkrho} we show the evolution of $K_{\rho i}$ as $U$
increases, for $t^{\prime }=2.5$. Clearly, for $U=10$ the
extrapolated $K_{\rho }$ are inconsistent and the system is in the
insulating phase. The metal-insulator transition seems to be near
$U/t=6$, where $K_{\rho }\sim 1/2 $. For lower values of $U$,
$K_{\rho }$ increases, which indicates that superconducting
correlations are the dominant ones inside the metallic phase. This
is in agreement with previous results.\cite{fab,kuro}

\section{The charge transition}
\begin{figure}[tbd]
\includegraphics[width=\columnwidth]{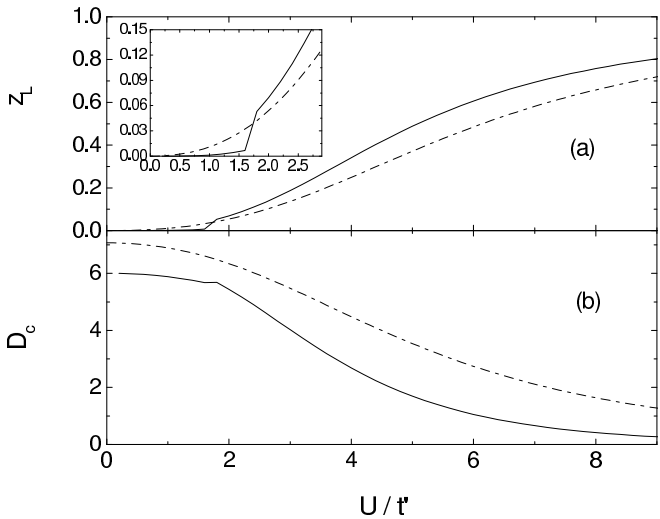}
\vskip -0.2cm \caption{\scriptsize Localization indicator $z_{L}$
and Drude weight $D_{c}$ as a function of $U$ for several system
sizes $L$ and $t^{\prime }=2.5$.  }\label{figdczl}
\end{figure}

In absence of disorder, the Drude weight $D_{c}$ defined in the
previous section is an indicator of the metal-insulator
transition. In particular, in the thermodynamic limit it should go
to zero in the insulating phase.\cite{kohn} In recent years, Resta
proposed to use as an alternative a localization indicator
$z_{L}$.\cite{resta} However, a simple symmetry argument shows
that this expression gives an incorrect result for periodic
systems with non integer number of particles per unit cell
$n.$\cite{ort1} One of us and Ortiz have generalized this concept
for correlated systems with rational $n$,\cite {ort1,ort2}
following a corresponding generalization for the charge Berry
phase.\cite{epl} In the present case, with $n=1$ one has
\begin{equation}
z_{L}=|\langle g|e^{i\frac{2\pi }{L}\sum_{j\sigma }jn_{j\sigma }}|g\rangle |.
\label{zl}
\end{equation}
and $z_{L}\rightarrow 0$ $(1)$ for a metal (insulator) in the
thermodynamic limit. Note that the operator entering Eq.
(\ref{zl}) is a displacement operator in momentum space. Then, for
an extended system with a well defined Fermi surface this operator
shifts the Fermi surface and the scalar product vanishes
($z_{L}=0$). Instead, for an uncorrelated band insulator the
operator reproduces the same state and $z_{L}=1$. Other cases with
either disorder or correlations were also discussed before.\cite
{ort1,ort2} One advantage of $z_{L}$ over $D_{c}$ is that it can
be calculated with DMRG,\cite{topo,ort2} although we are not
considering this here.

In Fig. \ref{figdczl} we show the results for $D_{c}$ and $z_{L}$
as a function of $U$, for $t^{\prime }=2.5$ and $L=8$,12. Both
quantities vary smoothly with $U$. While the results point towards
an insulating behavior for $U>4t^{\prime }=10$ (in particular
$z_{L}>0.5$ and increases with system size) and a metallic
behavior for small $U$ ($z_{L}\rightarrow 0$ and decreases with
$L$ for $U<1.75t^{\prime }$), it is not possible to determine
accurately the critical value of $U$ for which the transition
takes place. For large system sizes there is a level crossing in
the ground state, and while for one of the states $z_{L}$
increases with $L$, for the other one $z_{L}$ decreases with size.
This suggests to associate the crossing point with the
metal-insulator transition, which gives $U_{c}=1.75t^{\prime}$.

\begin{figure}[tbd]
\includegraphics[width=\columnwidth]{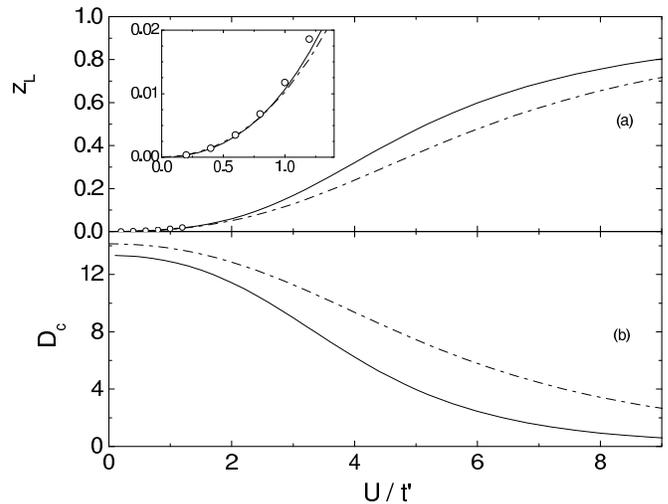}
\vskip -0.2cm \caption{\scriptsize Same as Fig. 4 for $t^{\prime
}=5$.}\label{figdczl2}
\end{figure}

As showed in Fig. \ref{figdczl2}, when $t^{\prime }$ is increased
to $t^{\prime }=5$ the values of $D_{c}$ increase and those of
$z_{L}$ decrease, suggesting an enlargement of the metallic
region. However, from the dependence on $L$ of the corresponding
quantities no safe conclusions can be drawn about the critical
value of $U$ at which the transition takes place. For the same
ratio $t^{\prime }=5$ we have also calculated numerically the
charge gap, including DMRG results up to 20 sites, using the
expression:
\begin{equation}
\Delta _{c}=E(L,N+1)+E(L,N-1)-2E(L,N).  \label{gapc}
\end{equation}

\begin{figure}[tbd]
\includegraphics[width=\columnwidth]{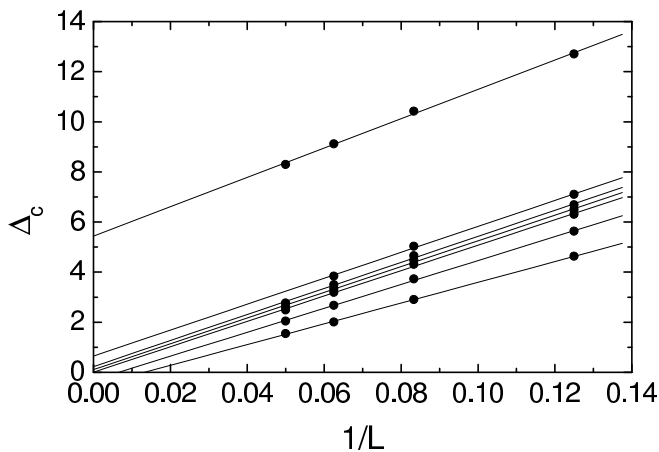}
\vskip -0.2cm \caption{\scriptsize Charge gap as a function of
$1/L$ for $t^{\prime }=5$ and different values of $U$. From bottom
to top: 2, 6, 8, 8.5, 9, 10, and 20. }\label{figgap}
\end{figure}

The results are shown in Fig. \ref{figgapc} together with a linear
extrapolation to the thermodynamic limit. Although a straight line
in $1/L$ fits well the results for all values of $U$, the
extrapolated $\Delta _{c}$ becomes negative for $U\leq 6$,
indicating again that finite size effects are important (since
$\Delta _{c}$ should be either zero or positive). Nevertheless,
for $U>10$ it seems clear that the gap is open.

\begin{figure}[tbd]
\includegraphics[width=\columnwidth]{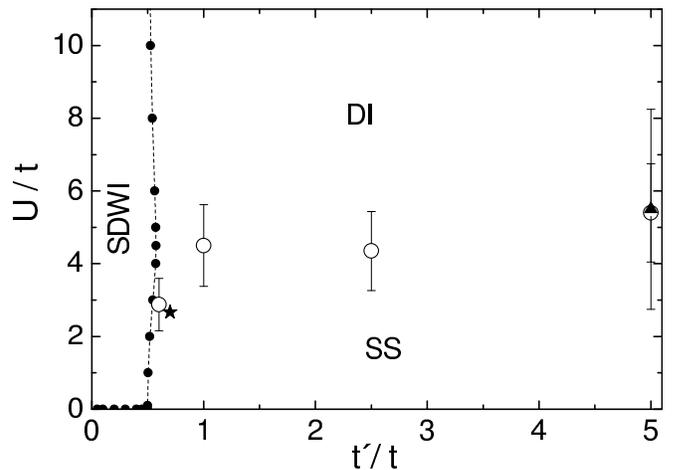}
\vskip -0.2cm \caption{\scriptsize Phase diagram of the
$t-t^{\prime }-U$ model. The star and triangle were taken from
Refs. \cite{aebi} and \cite{kuro}
respectively.}\label{diagfastpnuevo}
\end{figure}

Unfortunately, the charge Berry phase does not jump at the
metal-insulator transition, in contrast to the case of the Hubbard
model with correlated hopping.\cite{bos,topo} The study of this
transition using $K_{\rho i},D_{c},z_{L}$ and $\Delta _{c}$ as
explained above is not enough to lead to a well defined phase
boundary. With this information we can draw a tentative phase
diagram, presented in Fig. \ref{diagfastpnuevo}, with large error
bars for the metal-insulator DI-SS boundary. We also include some
points obtained by other authors using larger
systems.\cite{kuro,aebi} The boundary of the SDWI determined in
section II is also plotted.

\section{Summary and discussion}

We have studied the phase diagram of the $t-t^{\prime }-U$ model
at half filling, using different quantities calculated mainly by
numerical diagonalization of finite rings. Unfortunately, this
model has large finite size effects in the regime in which there
are four Fermi points in the non interacting case ($t^{\prime
}>0.5t$). This can be checked by calculating correlation functions
or other quantities as a function of size $L$ in the non
interacting limit, either for a ring or with open boundary
conditions. Kuroki \textit{et al.} argue that one should choose
system sizes for which the level structure of the different bands
in the non interacting case is similar.\cite{kuro}

The above mentioned difficulties prevented us to obtain
quantitatively valid results for the metal-insulator transition.
Nevertheless, our calculations of $K_{\rho }$ confirm the
dominance of superconducting correlations in the metallic phase
found earlier by analytical\cite{fab} and numerical\cite{kuro}
methods. Our tentative phase diagram also agrees with previous
studies.\cite{daul,kuro,aebi} For the spin transition, on the
contrary, our results show the $1/L^{2}$ scaling expected from
conformal field theory\cite{nom,nak} and, therefore, are reliable.
Moreover, they converge with high precision to the known result in
the strong coupling limit up to terms of order $1/U^{2}$. The jump
of the spin Berry phase and the equivalent level crossing method
have been successful in determining the opening of the spin gap in
different SU(2) invariant models,\cite{bos,nak,nak2,topo,pola,ih}
and its validity has been confirmed by other analytical and
numerical methods.\cite{bos,topo} In the present case, we became
aware of recent DMRG results for this transition based on
extrapolations of the dimer order parameter.\cite{tesis} These
results are very similar to ours except for a shift to larger $
t^{\prime }$ by about $\sim 0.7t$ for $U\sim 5t$. We believe that
this might be due to difficulties in detecting an exponentially
small dimerization order parameter or gap using DMRG.

An interesting question is what is the origin of the
superconducting correlations in this system, where the only
interaction is repulsive. An intuitive picture can be built
starting from the strong-coupling limit $U \rightarrow \infty $,
where the model is described by the spin Hamiltonian Eq.
(\ref{hheis}), whose properties are well known.\cite{nom} In the
dimerized phase, at the Majumdar-Ghosh point $J^{\prime}=0.5J$ the
ground state consists of nearest-neighbor spin singlets. These
electron pairs are static and the system is insulating. However,
when $U$ is reduced these pairs can overlap and become mobile,
behaving similarly to bosons with short range repulsion, for which
dominance of singlet superconducting correlations at large
distance can be expected.

\section*{Acknowledgments}

We thank Liliana Arrachea and Claudio Gazza for many useful
discussions. A.A.A. acknowledges the hospitality of ICTP, Trieste,
where part of this work was done, and MPI PKS, Dresden, for
computer time. This work was sponsored by PICTs 03-06343 and
03-03834 of ANPCyT. We are partially supported by CONICET.


\begin{thebibliography}{99}
\bibitem{note} At half filling, for $t^{\prime} < 0.5t$
bosonization leads to an Umklapp term of order $U$ that renders
the system insulating. For $t^{\prime} > 0.5t$, instead, there are
four Fermi points, and only effective Umklapp terms of order $U^2
/t$ survive.\cite{fab} Therefore, in this last case
the system is metallic even at half filling for small $U$.
\end{thebibliography}
\end{document}